\documentclass[aps,prl,twocolumn,showpacs,10pt]{revtex4-1}
\usepackage{amssymb}
\usepackage{amsmath}
\usepackage{epsfig}
\usepackage{color}
\usepackage[utf8]{inputenc}
\usepackage{graphicx}
\usepackage{enumerate}
\usepackage{array}
\usepackage[normalem]{ ulem }
\usepackage{soul}
\newcolumntype{M}[1]{>{\raggedright}m{#1}}
\newcommand{\be}{\begin{equation}}
\newcommand{\ee}{\end{equation}}

\begin{document}

\title{Phase separation of polymer-bound particles induced by loop-mediated 1D effective long-range interactions}

\author{G. David$^1$}
\author{J.-C. Walter$^1$}
\author{C. P. Broedersz$^2$}
\author{J. Dorignac$^1$}
\author{F. Geniet$^1$}
\author{A. Parmeggiani$^{1,3}$}
\author{N.-O. Walliser$^{1}$}
\author{J. Palmeri$^1$}
\email{email: john.palmeri@umontpellier.fr}
\affiliation{$^1$ Laboratoire Charles Coulomb (L2C), CNRS, Univ. Montpellier, Montpellier, France.}
\affiliation{$^2$ Arnold Sommerfeld Center for Theoretical Physics and Center for Nanoscience, Ludwig-Maximilian-Universität München, D-80333 München, Germany}
\affiliation{$^3$ DIMNP, CNRS, Univ. Montpellier, Montpellier, France.}

\date{\today}

\begin{abstract}
The cellular cytoplasm is organized into compartments. Phase separation is a simple manner to create membrane-less compartments in order to confine and localize particles like proteins. In many cases these particles are bound to fluctuating polymers like DNA or RNA. We propose a general theoretical framework for such polymer-bound particles and derive an effective 1D lattice gas model with both nearest-neighbor and emergent long-range interactions arising from looped configurations of the fluctuating polymer. We argue that 1D phase transitions exist in such systems for both Gaussian and self-avoiding polymers and, using a variational method that goes beyond mean-field theory, we obtain the complete mean occupation-temperature phase diagram.  To illustrate this model we apply it to the biologically relevant case of ParAB{\it S}, a prevalent bacterial DNA segregation system.
\end{abstract}

\maketitle

The confinement of chemical species, such as RNA or proteins, within the cytoplasm is mandatory for the spatio-temporal organization of chemical activities in the cell~\cite{Horwitz}. Cells compartmentalize the intracellular space using either membrane vesicles or membrane-less organelles. For the latter, cells may employ phase separation of chemical species in order to create localized high density regions in which specific reactions may occur~\cite{Hyman,Brangwynne13}. Such biological phase separation mechanisms often involve polymeric scaffolds like RNA or DNA to bind the chemical species~\cite{Brangwynne,Lee,Marko,Feric,Larson,Schumacher}.
A prominent example may be the formation of localized protein-DNA complexes during bacteria DNA segregation due to the {\it in vivo} ParAB{\it S} system~\cite{Broedersz,Sanchez,Legall,Walter2}. Although the molecular components of this widely conserved segregation machinery have been clearly identified, their dynamical interplay and the mechanism that leads to the condensation of the complexes remain elusive.\\
\indent More generally, despite extensive numerical studies~\cite{Broedersz,Johnson,Walter,Scolari}, it is still unclear theoretically how long 1D substrates like DNA polymers interact with particles to form 3D structures essential for the cellular cycle ~\cite{Brangwynne,Marko,Junier}. Interestingly, similar organizational principles may apply to the higher-order folding of chromatin and the interactions between topological domains in eukaryotic cells~\cite{Junier,Barbieri,Haddad,Jost}. A common theme is the mechanism of protein-induced polymer loop formation via bridging interactions and the role played by these loops in structuring DNA and creating localized protein-DNA complexes.
Three different basic models have been studied, mainly using simple mean-field Flory-type approaches and simulations: (i) sparse but fixed interacting sites~\cite{Junier,Scolari}, (ii) non-interacting mobile bound particles that can bind simultaneously to two polymer sites to form bridges~\cite{Barbieri,Scolari}, and (iii) mobile bound particles that can interact to form both nearest-neighbor (NN) and bridging bonds~\cite{Broedersz,Treut}.
\begin{figure}[t]
\includegraphics[scale=0.2]{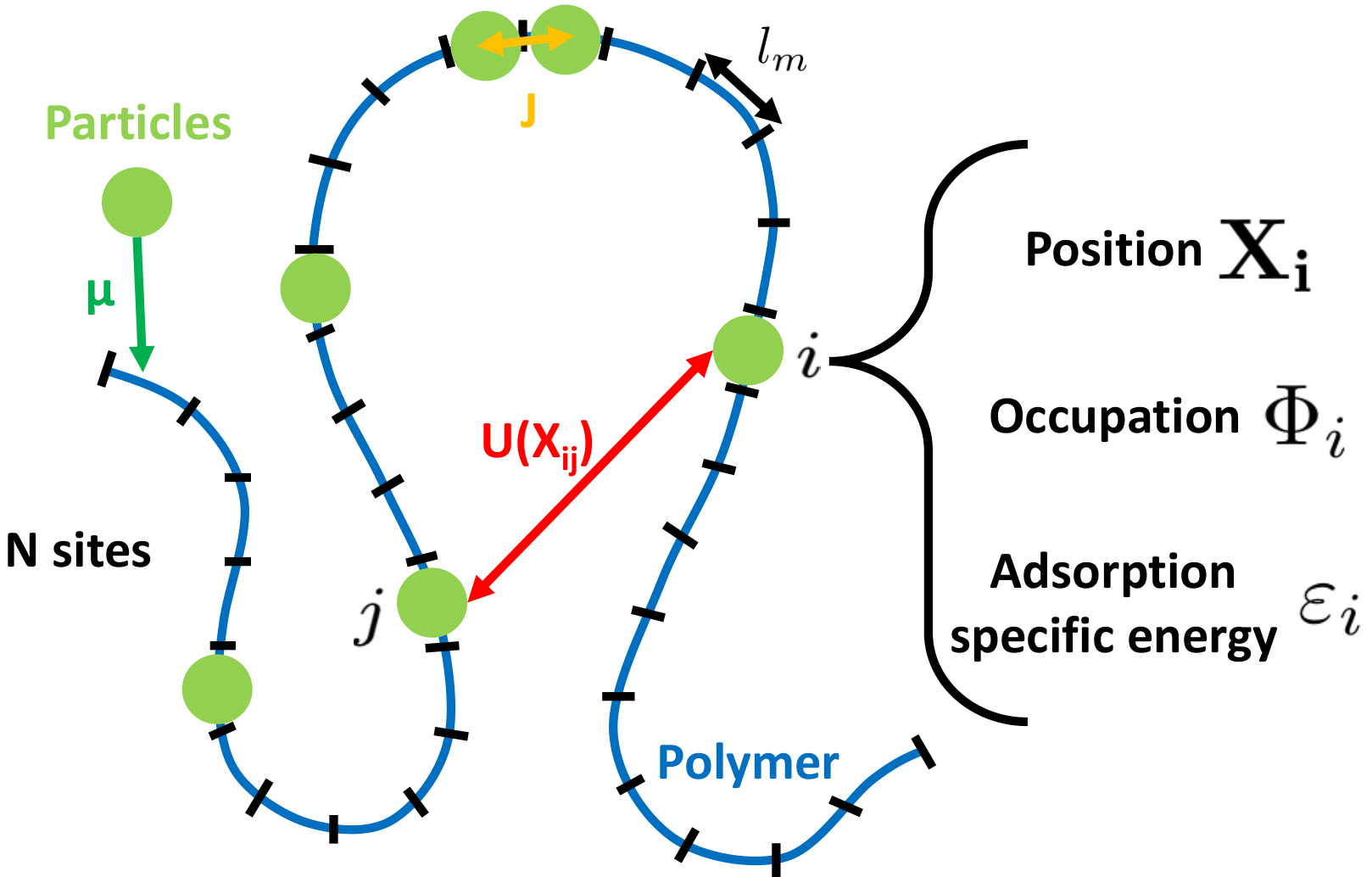}
   \caption{\label{Fig1} Schematic illustration of the coupled polymer-particle model. The polymer in 3D is divided into $N$ monomers, each having three attributes: a position vector $\mathbf{X}_i$, an occupation $\Phi_i$, and a local adsorption energy $\epsilon_i$. Loops form when particles far apart along the polymer interact at short range in 3D.}
\end{figure}
However, an analytical statistical mechanics framework is still needed to clarify the existence and nature of phase transitions in such systems. Here, we present an analytical Hamiltonian approach to case (iii) by introducing a basic microscopic particle-polymer model where all  relevant physical parameters appear explicitly. From this model, we derive an effective 1D lattice gas model with 1D temperature-dependent long-range interactions that arise from the 3D conformational fluctuations of the polymer. We show that the existence of a phase transition in this effective model depends on the exponent describing the asymptotic power law decay of the long-range interactions. We then propose a variational method that goes beyond mean-field theory (MFT) to compute the mean occupation-temperature phase diagram. Finally, for illustration, we apply our model to the bacterial partition system ParAB{\it S} and the formation of ParB{\it S} complexes. We propose a plausible explanation in terms of metastability for experiments showing the existence of high density ParB protein condensates only in the presence of specific binding sites. \\
\indent In our approach (see Fig.~\ref{Fig1}), the polymer consists of $N$ monomers (or sites) with each monomer capable of accommodating one bound particle. The effective monomer length $l_m$ corresponds to the footprint of one particle on the polymer, measured, for example, in terms of base pairs for DNA. Each site $i$ is characterized by its position in 3D space $\mathbf{X}_i$, its occupation $\Phi_i$ (equal to $1$ if a particle is bound and $0$ otherwise) and its on-site binding energy $\varepsilon_i$. This  energy allows us to implement local specific or non-specific binding. In the particle grand-canonical ensemble, the energy of a state $[\Phi_i ,{\bf{X}}_{i}]$ is
\begin{equation}
H[\Phi_i ,{\bf{X}}_{i}] = H_{\rm P}[{\bf{X}}_{i}] + H_{\rm SRLG}[\Phi_i]  + H_{\rm B}[\Phi_i ,{\bf{X}}_{i}].
\label{coupledmodel}
\end{equation}
The first term $H_{\rm P}[{\bf{X}}_{i}]$ describes the polymer configuration energy. The second is a 1D Short Range Lattice Gas (SRLG) Hamiltonian for bound particles,
\begin{equation}
\label{SRLG}
H_{\rm SRLG}[\Phi_i ] = - J \sum_{i =1}^{N - 1} \,   \Phi_{i + 1} \Phi_i  - \sum_{i = 1}^N  (\mu - \varepsilon_i) \,\Phi_i
\end{equation}
with NN spreading interaction coupling constant $J$ and chemical potential $\mu$. The contribution from 3D bridging interactions, giving the coupling between the bound particles and the fluctuating polymer, takes the form
\begin{equation}
H_{\rm B}[\Phi_i ,{\bf{X}}_{i} ] = \frac{1}{2} \sum_{i,j}^{N}\!{}^{'}   \Phi_i  U(X_{ij}) \Phi_j,
\end{equation}
with $X_{ij} = \vert {\bf{X}}_{i} - {\bf{X}}_{j} \vert$ and $U(X_{ij})$ the potential of 3D spatial interaction between particles. The prime on the sum means that $|i-j| \geq n_{\rm inf}$, where $n_{\rm inf}$ is the minimal internal distance in number of sites over which two particles can interact at long-range. \\
\indent
The polymer conformational  degrees of freedom can formally be integrated out, yielding a highly non-linear 1D effective free energy for the bound particles including two and all higher body interactions along the chain. Given the complexity of this coupled model,
we derive using a virial (cluster) expansion~\cite{Mayer, Feynman} a more amenable 1D effective model that retains only short and two body long-range interactions:
\begin{eqnarray}
\frac{\mathcal{Z}}{\mathcal{Z}_{\rm P} } & = &
\sum_{ \{ \Phi_i = 0,1\} }   e^{ -\beta \left(H_{\rm SRLG}[\Phi_i] - \beta^{-1} \ln   \langle e^{-\beta H_{\rm B}[\Phi_i ,{\bf{X}}_{i}]} \rangle_{\rm P} \right) } \nonumber \\
& \approx & \sum_{ \{ \Phi_i = 0,1\} }   e^{ -\beta \mathcal{F}_{\rm LRLG}[\Phi_i] }
\end{eqnarray}
where $\beta = 1/(k_{\rm B}T)$, $\langle \cdot \rangle_{\rm P}$ denotes an average over polymer conformations, $\mathcal{Z}_{\rm P}$ is the partition function of the bare polymer, and $\mathcal{F}_{\rm LRLG}[\Phi_i]$ is a 1D long-range Lattice Gas (LRLG) effective (temperature dependent) free energy:
\begin{equation}
\mathcal{F}_{\rm LRLG}[\Phi_i] = H_{\rm SRLG}[\Phi_i ] - \frac{1}{2} \sum_{i,j}^{N}\!{}^{'} \Phi_i G_{ij} \Phi_j
\label{FLRLG}
\end{equation}
The second term of Eq.~\eqref{FLRLG} is an effective 1D long-range bridging interaction between particles on the polymer that depends on the distance along the chain and arises after the chain conformational fluctuations have been integrated out, giving rise to the temperature dependence of $\mathcal{F}_{\rm LRLG}$. The kernel,
\begin{equation}
G_{ij} = 4\pi \beta^{-1} \int_0^\infty dR \, R^2  \, \left[ e^{-\beta U(R)} -1 \right] \, P_{ij}(R),
\label{kernel}
\end{equation}
is obtained by performing a generalized virial expansion (assuming isotropy) with
\begin{equation}
P_{ij}(R) = \langle \delta (R - \vert {\bf{X}}_i- {\bf{X}}_j \vert) \rangle_{\rm P}
\end{equation}
the polymer distribution function. The effective free energy $\mathcal{F}_{\rm LRLG}$ is therefore completely defined by the polymer and particle parameters.
The effective long-range interaction encoded by the kernel $G_{ij}$ implicitly sums over all possible loops formed by the polymer segment bounded by the two bridging particles.
This approach accounts exactly for two-body interactions and should therefore be valid for sufficiently low polymer monomer 3D spatial density (as in Flory-type approximations~\cite{DeGennes1,DeGennes,Flory}). There will be no restriction, however, on the 1D occupation along the polymer.
\\
\indent The possibility that the LRLG model exhibit a phase separation, while the 1D SRLG model does not, is thus completely dependent on the asymptotic behavior $\vert i - j \vert \rightarrow \infty$ of the kernel $G_{ij}$. The asymptotic behavior of $P_{ij}(R)$ is~\cite{DeGennes}
\begin{equation}
P_{ij}(R) \underset{\frac{R}{R_{ij}} \to 0}{\longrightarrow} \frac{c_0}{R_{ij}^3} \left( \frac{R}{R_{ij}} \right)^g ,
\label{scalerelationpij}
\end{equation}
where $c_0$ is a constant and $R_{ij} = \langle  X_{ij}^2 \rangle_{\rm P}^{1/2} = b \vert i - j \vert^{\nu}$ is the root-mean-square monomer $i$-to-$j$ distance with $b$ the Kuhn length.
The exponents $\nu$ and $g$ depend on the chosen polymer statistics. In the absence of the polymer, the monomers form an ideal gas and $P_{ij}(R)$ is replaced by the inverse system volume $V^{-1}$ in Eq.~\eqref{kernel}. The above approach then reduces to the usual non-ideal gas virial expansion.
By contrast, particle-particle correlations arise from the polymer connectivity due to the presence of $P_{ij}(R)$ in the kernel $G_{ij}$. Bound particles closer on the chain thus experience enhanced two-body interactions down to a lower limit imposed by polymer rigidity and self-avoidance. \\
\indent By inserting Eq.~\eqref{scalerelationpij} in~\eqref{kernel}, we obtain the asymptotic behavior of the long-range interaction,
$G_{ij} \sim \vert i - j \vert^{-\alpha}$
with $\alpha = (3+g)\nu$. The effective 1D LRLG model clearly falls into the universality class of the well known 1D long-range Ising model (LRIM)~\cite{Luijten}, aside from an additional NN interaction that also appears in the effective inverse square LRIM approach to the Kondo problem~\cite{PWA}. The exponent $\alpha$ is the key parameter to predict phase transitions in the LRIM~\cite{Dyson}. Ferromagnetic-like phase transitions occur for a positive kernel and $1 < \alpha < 2$ (Dyson criterion) and critical exponents are classical for $1 < \alpha < 3/2$~\cite{Mukamel}. The case $\alpha = 2$ leads to the 1D analog of the  Berezinky-Kosterlitz-Thouless phase transition~\cite{PWA,Kosterlitz}. \\
\indent
Interestingly, the Dyson criterion depends here only on the polymer properties and
it is straightforward to obtain the values of $\alpha$ for the Gaussian and self-avoiding polymer (SAP) distributions.
For a Gaussian polymer $\nu = 1/2$ and $g =0$, and therefore $\alpha = 3/2$. For a SAP $\alpha \approx 1.92$, since $\nu \approx 0.588$ and $g \approx 0.27$~\cite{DeGennes}. Therefore, the Dyson criterion for $\alpha$ is fulfilled and these two polymer models are expected to lead to phase separation. For an infinite compact globular polymer, we expect Gaussian behavior for interior monomers owing to internal screening of polymer self-avoidance~\cite{schbio}. Typical polymer conformational statistics therefore lead to a LR interaction decay exponent $\alpha$ that ensures the existence of a 1D phase transition for bound particles. \\
\indent Using a variational method~\cite{Feynman}, we proceed by finding the coexistence and spinodal curves to construct the entire LRLG phase diagram. Assuming homogeneous non-specific binding, a constant $\epsilon_i$ can be absorbed into the definition of the chemical potential, and we rewrite the free energy $\mathcal{F}_{\rm LRLG}$ as the sum of two parts by introducing a variational parameter $\mu_{\rm 0}$:
\begin{equation}
\mathcal{F}_{\rm LRLG}[\Phi_i] = H_{\rm 0} + \Delta H,
\label{vardivision}
\end{equation}
where
\begin{equation}
H_{\rm 0} = - J \sum_{i =1}^{N - 1} \,   \Phi_{i + 1} \Phi_i  - \mu_{\rm 0} \sum_{i = 1}^N \,\Phi_i
\end{equation}
and
\begin{equation}
\Delta H = -\frac{1}{2} \sum_{i,j}^{N}\!{}^{'} \Phi_i G_{ij} \Phi_j - (\mu - \mu_{\rm 0}) \sum_{i = 1}^N  \,\Phi_i.
\label{DH}
\end{equation}
$H_0$ is just the Hamiltonian of another 1D SRLG (see Eq.~\eqref{SRLG}) with an effective chemical potential $\mu_{\rm 0}$ and therefore has the advantage of being exactly solvable. For $J = 0$, the variational method is equivalent to the MFT one, which consists in moving the NN interaction (term in $J$) from $H_{\rm 0}$ to $\Delta H$ (see Supplementary Material~\cite{SMref}).  MFT, which incorrectly predicts a 1D phase in the absence of bridging, is improved by the optimal choice for $\mu_0$ when $J > 0$, because correlation effects, missed entirely by MFT, are approximately accounted for in the variational $H_{\rm 0}$. This variational method is exact for the infinite range lattice gas (or Ising model~\cite{Kijewski,Petrosyan}) and therefore we expect it to lead to reasonably accurate results for the LRLG.
The division in Eq.~\ref{vardivision} leads to a trial grand potential $\Omega_{\rm V} = \Omega_{\rm 0} +  \langle \Delta H \rangle_{\rm 0} \ge \Omega_{\rm LRLG}$,
where $\Omega_{\rm 0}$ is the grand potential related to $H_{\rm 0}$ and $\langle \cdot \rangle_{\rm 0}$ denotes an average with respect to $H_{\rm 0}$. In the thermodynamic limit ($N \rightarrow \infty$), $\Omega_{\rm 0} = - N k_{\rm B} T \ln \lambda_+$,
where $\lambda_+$ is the largest of the two eigenvalues $\lambda_\pm$ which arise from the transfer matrix method applied to the SRLG model~\cite{Mccoy}:
\begin{equation}
\lambda_{\pm} = e^{Y} \left[ \cosh(Y) \pm \sqrt{\sinh^2 (Y) + e^{-\beta J} } \right],
\end{equation}
where $Y = \beta ( J + \mu_{\rm 0} )/2$. The second term in $\Omega_{\rm V}$,
\begin{equation}
\langle \Delta H \rangle_{\rm 0} = \frac{1}{2} \sum_{i,j}^{N}\!{}^{'} \langle \Phi_i  \Phi_j \rangle_{\rm 0} - ( \mu - \mu_{\rm 0} ) \sum_{i = 1}^N  \,\langle \Phi_i \rangle_{\rm 0},
\end{equation}
involves the mean occupation in the ensemble $H_{\rm 0}$, $\Phi_{\rm 0}  \equiv \langle \Phi_i \rangle_{\rm 0} $, where
\begin{equation}
\langle \Phi_i \rangle_{\rm 0} =  -\frac{1}{N} \frac{\partial \Omega_{\rm 0}}{\partial \mu_{\rm 0}}
 =  \frac{1}{2} \left( 1 + \frac{\sinh(Y)}{\sqrt{\sinh^2(Y) + e^{-\beta J}}} \right),
\label{Phi0}
\end{equation}
and the two-site correlation function,
\begin{equation}
\langle \Phi_i  \Phi_j \rangle_{\rm 0} =  \Phi_{\rm 0}^2 + \Phi_{\rm 0} \left( 1 - \Phi_{\rm 0} \right) e^{-\vert i - j \vert/\xi_{\rm LG}},
\end{equation}
in the thermodynamic limit with $\xi_{\rm LG} = -1/\ln r_{\rm LG}$ the SRLG correlation length and $r_{\rm LG} \equiv {\lambda_-}/{\lambda_+}$.
The optimization equation $\left( \partial \Omega_{\rm V}/\partial \mu_{\rm 0} \right)_{\mu_{\rm 0} = \mu_{\rm 0}^\star} = 0$ gives the optimal value $\mu_{\rm 0}^\star$ of $\mu_{\rm 0}$:
\begin{eqnarray}
\mu - \mu_0^\star & = & 2 \Phi_{\rm 0}^\star \left[ S' - S \right] - S' \nonumber \\
& - & \Phi_{\rm 0}^\star (1-\Phi_{\rm 0}^\star) ( 1 - 2\Phi_{\rm 0}^\star ) S'' \beta
\left( \frac{\partial \Phi_{\rm 0}}{ \partial \mu_{\rm 0}} \right)_{\mu_{\rm 0} =\mu_{\rm 0}^\star}^{-1}
\end{eqnarray}
with $\Phi_{\rm 0}^\star = \Phi_{\rm 0}(\mu_{\rm 0}^\star)$ and where the sums $S$, $S'$ and $S''$, defined as
$S = \sum_{k = n_{\rm inf}}^{\infty} G_k$, $S' =  \sum_{k = n_{\rm inf}}^{\infty} G_k  r_{\rm LG}^k$, and $S'' =  \sum_{k = n_{\rm inf}}^{\infty} G_k  k \, r_{\rm LG}^k$, depend crucially on the long-range behavior of the kernel $G_{ij} = G_{i-j}$ (see~\cite{SMref}). The best variational approximation to the exact grand potential $\Omega_{\rm LRLG}$ is the optimal grand potential, $\Omega_{\rm V}^\star = \Omega_{\rm V}(\mu_{\rm 0}^\star)$, from which we obtain the average site occupation $\Phi \equiv -N^{-1} \partial \Omega_{\rm V}^\star / \partial \mu$. This last definition, along with the optimization condition, leads to $\Phi = \Phi_{\rm 0}^\star$ and since
Eq.~\eqref{Phi0} can be inverted to obtain $\mu_{\rm 0}^\star$ in terms of $\Phi_{\rm 0}^\star$, it is possible to write $\Omega_{\rm V}^\star$ entirely in terms of $\Phi$ (see~\cite{SMref}):
\begin{eqnarray}
\frac{\Omega_{\rm V}^\star}{N}
 & = &  \frac{\Omega_0(\Phi)}{N} + \Phi^2 \left( S - S' \right) \nonumber \\
& + &   \Phi^2 (1-\Phi) ( 1 - 2\Phi ) \beta S'' \left( \frac{\partial \Phi_0}{ \partial \mu_0} \right)_{\mu_0 =\mu_{\rm 0}^\star}^{-1}.
\end{eqnarray}
We therefore obtain analytical variational expressions for the chemical potential $\mu$, the LRLG pressure $P \approx -\Omega_{\rm V}^\star/(N l_m) $ as functions of $\Phi$ that can be used to obtain the coexistence and spinodal curves~\cite{Huang} (see~\cite{SMref}). \\
\indent For simplicity, we illustrate our results for the case of an attractive square well (SW) particle interaction of depth $u_0$, range $a$ and hard core $\sigma$~\cite{Junier, Scolari}. The asymptotic long distance behavior (for $R_{ij}/b \gg 1$)  is therefore given by
$G_{ij} \underset{|i-j| \rightarrow \infty}{\longrightarrow} K_{\rm SW}|i-j|^{-\alpha}$
where
\begin{eqnarray}
K_{\rm SW} &=& 4\pi \beta^{-1}
\frac{c_0}{3+g} \left( \frac{\sigma}{b} \right)^{3+g} \nonumber\\
& & \times \left\{
\left( e^{\beta u_0} -1 \right) \left[ \left( \frac{a}{\sigma}\right)^{3+g} - 1 \right]
-1
\right\}.
\end{eqnarray}
This result allows us to illustrate generic behavior for potentials with short range repulsion and longer range attraction: $K_{\rm SW}$ is positive (attractive) at low enough $T$ and decreases monotonically with decreasing slope for increasing temperature, eventually becoming negative (repulsive) at high enough $T$ due to short range repulsion. In the attractive regime of interest, $K_{\rm SW}$ increases with $u_0$ and $a$ and decreases with the Kuhn length $b$, $\sigma$, and polymer exponent $g$ because chain stiffness and polymer self-avoidance inhibit particle-particle bridging. \\
\indent
We apply our LRLG  model with the SW potential to study  phase separation in the ParAB{\it S} partition system. This molecular machinery is composed of three components: a DNA sequence {\it parS}, and two protein species ParB and ParA. We focus on one of its key elements: the formation of ParB  aggregates around {\it parS}. ParB proteins can bind to DNA non-specifically and specifically on the {\it parS} sequence~\cite{Rodionov}. Once bound to DNA, ParB proteins can mutually interact through both spreading and bridging interactions (see Fig.~\ref{Fig1}), which lead to the formation of ParB{\it S} partition complexes~\cite{Broedersz,Graham}. Although we now have a better understanding of segregation dynamics~\cite{Walter2}, the conditions of complex formation are still poorly understood.\\
\indent
With our model we are now positioned to investigate whether or not the formation of ParB{\it S} complexes could be the result of a 1D phase separation between states of high and low ParB occupation on the DNA, qualitatively similar to conventional liquid-vapor phase separation. The available data for ParB allow us to parameterize the LRLG model at room temperature $T_{\rm r} = 300~\rm{K}$ (See Fig.~\ref{Fig2} and~\cite{SMref}).
\begin{figure}[t]
\includegraphics[scale=0.32]{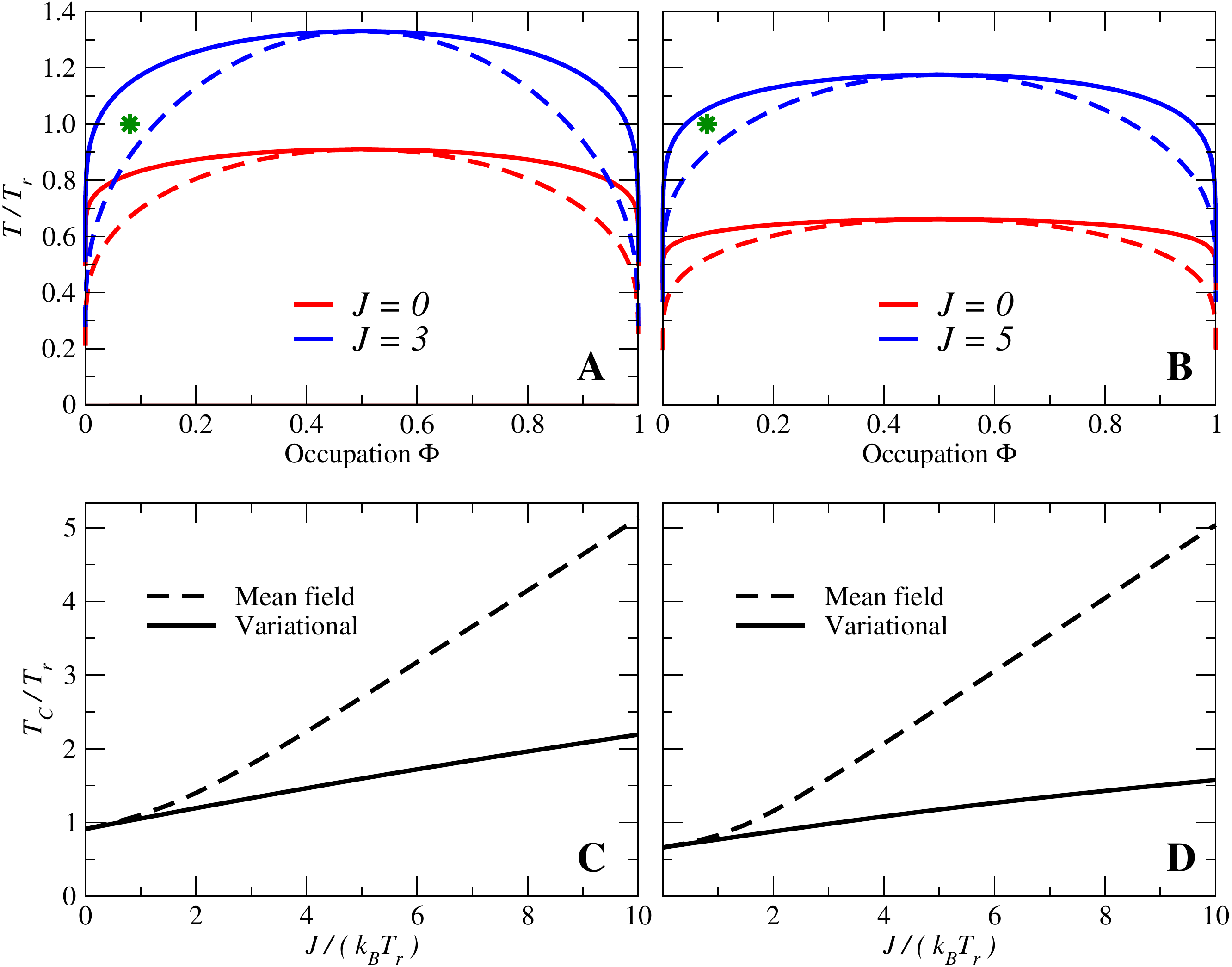}
   \caption{\label{Fig2} Phase diagrams for polymer-bound particles. Model parameters (see~\cite{SMref}): $l_m = 5.44~\rm{nm}$, $\sigma = l_m$,  $b = 23.6$~nm,  $n_{\rm inf} = 10$, $a = 2\sigma$, and $u_0 = 3~k_{\rm B} T_{\rm r}$. Green star: biological conditions for the bacterial F-plasmid ($\Phi = 0.08$ at room temperature $T_{\rm r}$). \textbf{(A)}: Gaussian polymer. Solid (dotted) line represents the coexistence (spinodal) curve for $J=0$  (red) and $J=3~k_{\rm B} T_{\rm r}$  (blue).  \textbf{(B)}: Self avoiding polymer (SAP) with $J=0$  (red) and $J=5~k_{\rm B} T_{\rm r}$  (blue). \textbf{(C)}:  Critical temperature $T_{\rm c}$  for the Gaussian polymer: variational approach (solid line) and MFT (dotted line).
   \textbf{(D)}: Same as \textbf{(C)}, but for the SAP.}
\end{figure}
Figures~\ref{Fig2}AB show the phase diagrams obtained using Gaussian polymer or SAP statistics. The coexistence and spinodal curves are obtained from the equality of pressure and chemical potential in the two phases and the divergence of the isothermal compressibility, respectively. The critical temperature is found in the limit $\Phi \rightarrow \Phi_{\rm c} =1/2$ (see~\cite{SMref}). This leads to the variational critical temperature as a solution to the following implicit equation:
\begin{equation}
\frac{ T_{\rm c}^{\rm V} }{ T_{\rm r} }\
= \frac{1}{2 k_B T_{\rm r}}
\left[ (S_{\rm c}-S_{\rm c}')
\exp
\left(
\frac{J}{2 k_B T_{\rm c}^{\rm V}}
\right) - S_{\rm c}''
\right],
 \label{tcv}
\end{equation}
where the subscript c indicates quantities evaluated at the critical point.
We observe that $T_{\rm c}^{\rm V}$ grows with $J$ (Figs.~\ref{Fig2}CD) and that this effect is severely overestimated by MFT,
for which (see~\cite{SMref})
\be
\frac{ T_{\rm c}^{\rm MFT} }{ T_{\rm r} } =
\frac{1}{2 k_{\rm B} T_{\rm r} }
\left[
J + S(T_{\rm c}^{\rm MFT})
\right].
\label{tcmft}
\ee
In the asymptotic kernel approximation adopted here
\be
S(T) = \sum_{k = n_{\rm inf}}^{\infty}\!\!\!\! G_k \approx K_{\rm SW} (T)
\left[
\zeta(\alpha) - \sum_{k=1}^{n_{\rm inf}-1} \frac{1}{k^\alpha}
\right]
\label{st}
\ee
with $\zeta$ the Riemann zeta function.
A simple approximation based on the weak temperature dependence of $K_{\rm SW}(T)$ for $T > T_{\rm r}$ and obtained by evaluating $S$ in Eq.~\eqref{tcmft} at $T_{\rm r}$ explains the linear dependence of $T_{\rm c}^{\rm MFT}$ on $J$ for large $J$ (see~\cite{SMref}). The temperature dependence of the kernel is, however, crucial in determining the critical temperature for small $J$.
The variational result for the critical temperature is also  close to being linear in $J$ for large $J$ and heuristically can be obtained from MFT by evaluating evaluating $S$ at $T_{\rm r}$ and replacing $J$  by $J/3$.

The expression~\eqref{st} indicates how the critical temperature is crucially determined by $n_{\rm inf}$, the polymer persistence length in site number, by reducing the weight of the LR interaction contribution~\cite{Marko,Walter}. In Fig.~\ref{Fig2}, the lower $T_{\rm c}$ shown by the SAP compared with the Gaussian polymer at constant $J$ is due to the faster decay of the LR interaction (larger $\alpha$), despite the larger value of the SAP $K_{\rm SW}$ (see~\cite{SMref}). $T_{\rm c}$ is non-zero even for $J = 0$, but is far below room temperature. Therefore, the system does not exhibit phase separation without spreading interactions at this temperature. Both short range spreading with reasonable biological values for $J$ ($\sim 3$-$6~k_{\rm B} T_{\rm r}$) and long-range bridging interactions are thus required at room temperature to form ParB condensates in our model, as suggested by Monte Carlo simulations~\cite{Broedersz} and experiments~\cite{Diaz,Graham}.\\
\indent The ParAB{\it S} system ensures the segregation in {\it E. coli} of relatively short circular DNA strands
called F-plasmids. For an F-plasmid of linear size $\sim 60~\rm{kbp}$ and an average number of $300$ ParB~\cite{Bouet}, the mean occupation is $\Phi \approx 0.08$. Its position in the phase diagram (green star in Fig.~\ref{Fig2}AB) shows that for reasonable values of $J$ the system may exist in the low occupation metastable coexistence region at room temperature, providing a plausible explanation for the experimental observations~\cite{Sanchez}: without the {\it parS} sequence, experiments show a homogeneous ParB distribution in the cell, while with {\it parS} a ParB{\it S} complex forms. Thus, {\it parS} could provide the energy required to overcome the nucleation barrier and allow the system to switch from the metastable homogeneous state to the stable coexistence phase, in which ParB proteins form a stable cluster on the DNA around {\it parS}. Experimentally, this system should follow the conventional behavior of liquid-vapor phase transitions:
(i) in the low occupation metastable region, the system can form relatively high density ParB{\it S} complexes with only a small total number of intracellular proteins, and (ii)
ParB over- or under-expression will favor or repress the formation of ParB{\it S} complexes depending on the position in the phase diagram. Indeed, systems without {\it parS} but with sufficiently high ParB occupation would be in the unstable coexistence area and should therefore form protein (\textit{liquid}) droplets spontaneously in a low occupation (\textit{vapor}) background, the homogeneous state being unstable in this case. On the contrary, systems with too few ParB proteins would be in the low occupation \textit{vapor} region, losing the ability to form complexes even in the presence of {\it parS}. \\
\indent In this article, we proposed a general theoretical framework for the physics of particles interacting on a polymer fluctuating in 3D that leads naturally to an effective 1D LRLG model. We established a criterion for the existence of a 1D phase transition based on the exponent $\alpha$ controlling the asymptotic decay of the LR interactions, which depends only on the polymer exponents $\nu$ and $g$. Since this criterion is satisfied for standard polymer models, the conformational fluctuations of linear structures like DNA produce effective 1D long-range interactions between bound particles that lead to 1D particle phase separation along the polymer. We used our theoretical approach to construct the whole phase diagram of the ParB{\it S} bacterial DNA segregation system and concluded that the formation of ParB{\it S} complexes results from activated phase separation in the low ParB occupation metastable region. This general mechanism for triggering the formation of polymer-bound protein complexes via small nucleation sites may play an important role in membrane-less cell compartmentalization. \\
\indent Our method may also be used to derive the 1D particle distribution along the polymer and the 3D particle density of the condensate that forms around a specific binding site, both of which are accessible experimentally~\cite{Sanchez,Diaz}.  Finally, to facilitate quantitative testing of the present model, it would also be of great interest to find an \textit{in vitro} biomimetic system of interacting polymer-bound particles that could be studied experimentally.

\begin{acknowledgments}
This project received partial financial support from the French Agence Nationale
de la Recherche (\textit{Imaging and Modeling Bacterial Mitosis} project ANR-14-CE09-0025-01),  the CNRS D\'efi Inphyniti (\textit{Projet Structurant} 2015-2016), and the program ‘Investissements d’Avenir’ ANR-10-LABX-0020 and Labex NUMEV (AAP 2013-2-005, 2015-2-055, 2016-1-024) (GD, JCW, JD, FG, AP, NOW, JP). GD acknowledges doctoral thesis support from the French \textit{Minist\`ere de l’enseignement supérieur}.
This project was also supported in part by the German Excellence Initiative via the program NanoSystems Initiative Munich (NIM) (CPB), the Deutsche Forschungsgemeinschaft (DFG) Grant TRR174 (CPB). We would like to thank J-Y Bouet, M. Nollman, and N. Wingreen for interesting discussions on the ParAB{\it S} system.
\end{acknowledgments}

\end{document}